# Optical spectroscopic investigation on the temperature-dependent electronic structure evolution of the $J_{\text{eff},1/2}$ Mott insulator $Sr_2IrO_4$

S. J. Moon,[1] Hosub Jin,[2] W. S. Choi,[1] J. S. Lee,[1,*] S. S. A. Seo,[1,†] J. Yu,[2] G. Cao,[3] T. W. Noh,[1] and Y. S. Lee[4,‡]

[1] *ReCOE & FPRD, Department of Physics and Astronomy, Seoul National University, Seoul 151-747, Korea*

[2] *CSCMR & FPRD, Department of Physics and Astronomy, Seoul National University, Seoul 151-747, Korea*

[3] *Department of Physics and Astronomy, University of Kentucky, Lexington, Kentucky 40506, USA*

[4] *Department of Physics, Soongsil University, Seoul 156-743, Korea*

We investigated the temperature-dependent evolution of the electronic structure of the $J_{\text{eff},1/2}$ Mott insulator $Sr_2IrO_4$ using optical spectroscopy. The optical conductivity spectra $\sigma(\omega)$ of this compound has recently been found to exhibit two *d-d* transitions associated with the transition between the $J_{\text{eff},1/2}$ and $J_{\text{eff},3/2}$ bands due to the cooperation of the electron correlation and spin-orbit coupling. As the temperature increases, the two peaks show significant changes resulting in a decrease in the Mott gap. The experimental observations are compared with the results of first-principles calculation in consideration of increasing bandwidth. We discuss the effect of the temperature change on the electronic structure of $Sr_2IrO_4$ in terms of local lattice distortion, excitonic effect, electron-phonon coupling, and magnetic ordering.

## I. INTRODUCTION

Transition metal oxides (TMOs) have been the topic of much recent research due to their numerous intriguing phenomena, such as Mott transition, superconductivity, colossal magnetoresistance, and spin/orbital ordering.[1-3] The large variety of phenomena originates from various interactions, including on-site Coulomb, hopping, electron-phonon, and spin-orbit, which mediate the coupling among spin, charge, orbital, and lattice degrees of freedom. Especially, the spin-orbit coupling (SOC) is an important consideration in describing the physical properties of 5$d$ TMOs. While the electron-electron interaction becomes weaker, the SOC becomes about 0.4-0.5 eV for 5$d$ TMOs, which is much larger than those of 3$d$/4$d$ TMOs.[4] As the SOC energy scale is comparable to those of other fundamental interactions, it could give rise to some interesting properties of 5$d$ TMOs.

Sr$_2$IrO$_4$ has a K$_2$NiF$_4$-type layered perovskite structure with five valence electrons in the 5$d$ $t_{2g}$ states. As the 5$d$ orbitals are spatially more extended than 3$d$ and 4$d$ orbitals, their bandwidth (Coulomb interactions) could be larger (smaller) than those of 3$d$ and 4$d$ orbitals. Thus 5$d$ TMOs are expected to have a metallic ground state. Contrary to the conventional understanding of the electron correlation, Sr$_2$IrO$_4$ was found to be an insulator with weak ferromagnetic ordering below $T_M$=240 K.[5, 6] It has recently been found that the peculiar insulating state could be due to the cooperation of the SOC and on-site Coulomb interaction, $U$, and that the electronic ground state should be the $J_{\text{eff},1/2}$ Mott state.[7, 8]

The $J_{\text{eff},1/2}$ state can be expressed as

$$\left| J_{\text{eff}} = 1/2, m_{J_{\text{eff}}} = \pm 1/2 \right\rangle = \frac{1}{\sqrt{3}} \left( \pm \left| xy \right\rangle \left| \pm \right\rangle + \left| yz \right\rangle \left| \mp \right\rangle \pm i \left| zx \right\rangle \left| \mp \right\rangle \right), \quad (1)$$

where $\left| + \right\rangle$ and $\left| - \right\rangle$ represent spin-up and spin-down states, respectively. The $J_{\text{eff},1/2}$ state is strikingly different from the $d$-orbital states of the 3$d$ or 4$d$ TMOs. In 3$d$ or 4$d$ TMOs, in which the SOC is small, the electronic state is determined by the change in the crystal field and the resulting electronic state is described by real wavefunctions in the forms of the $t_{2g}$ ($xy$, $yz$, and $zx$) and $e_g$ ($x^2$-$y^2$ and $3z^2$-$r^2$) orbitals.[2, 9, 10] However, as shown in the Eq. (1), the SOC-induced $J_{\text{eff},1/2}$ state is the complex state composed of the $xy$, $yz$, and $zx$ orbitals and the spins. This state can be a new basis for interesting emergent phenomena, such as the low-energy Hamiltonian of the Kitaev model[11] and the room-temperature quantum spin Hall effect in a honeycomb lattice.[12] Therefore, investigating the characteristic properties of the $J_{\text{eff},1/2}$ state and establishing a new territory of the strongly correlated electron systems are important.

In this paper, we investigated the temperature-dependent optical conductivity spectra $\sigma(\omega)$ of $J_{\text{eff},1/2}$ Mott insulator Sr$_2$IrO$_4$. The two-peak structure of the $d$-$d$ transition in $\sigma(\omega)$, which is associated with the Hubbard bands of the $J_{\text{eff},1/2}$ state and the $J_{\text{eff},3/2}$ bands, showed significant temperature dependence. Spectral weight (SW) redistribution occurred between the two optical transitions, and the Mott gap drastically decreased with increasing temperature. We performed a first-principles calculation and found that our experimental observation could be associated with the change in the bandwidth of the $J_{\text{eff}}$ states. We discussed possible effects of the magnetic ordering and electron-phonon coupling to the unusual temperature dependence of $\sigma(\omega)$.

## II. EXPERIMENTAL AND THEORETICAL METHODS

Single crystals of $Sr_2IrO_4$ were grown using flux technique.[5] We measured near-normal incident *ab* plane reflectance spectra $R(\omega)$ in the energy region between 5 meV and 6 eV as a function of temperature, and between 6 and 20 eV at room temperature. In the temperature range between 10 K and 300 K, a continuous-flow liquid-helium cryostat was used. For higher temperature measurements up to 500 K, a custom-built sample holder with a heating system was used.[13] We used a Fourier transform spectrometer (Bruker IFS66v/S) and a grating-type spectrophotometer (CARY 5G) in the photon energy region of 0.1-1.2 eV and 0.4-6.0 eV, respectively. In the deep ultraviolet region above 6.0 eV, we used synchrotron radiation from the normal incidence monochromator beamline at Pohang Light Source. The corresponding optical conductivity spectra $\sigma(\omega)$ were obtained using the Kramers-Kronig transformation.[14]

The density functional theory code OpenMX was used for the theoretical calculation on the electronic structure. The calculation is based on a linear combination of pseudo-atomic orbitals method, whereby both the local density approximation+$U$ method and SOC (LDA+$U$+SOC) are included via a relativistic *J*-dependent pseudo-potential scheme in the non-collinear density functional formalism. Details of the calculation method can be found in Ref. 15.

## III. RESULTS AND DISCUSSION

A. Temperature dependence of $\sigma(\omega)$

Figure 1(b) shows $\sigma(\omega)$ of $Sr_2IrO_4$ at various temperatures. The sharp spikes below 0.1 eV are due to the optical phonon modes. The spectra show two optical transitions above 0.2 eV, labeled as $\alpha$ and $\beta$ in Fig. 1(b). As the charge transfer transition from O 2*p* to Ir 5*d* orbital states are located above 3 eV, the peaks $\alpha$ and $\beta$ should be *d-d* transitions.[6] This two-peak structure is associated with the SOC-induced Mott insulating state.[8] The strong SOC of 5*d* Ir ion splits the $t_{2g}$ orbital states into the total angular momentum $J_{\text{eff},1/2}$ and $J_{\text{eff},3/2}$ states. The $J_{\text{eff},1/2}$ bands are partially occupied by a single electron and the $J_{\text{eff},3/2}$ bands are fully occupied by four electrons. The effect of $U$ splits the partially filled $J_{\text{eff},1/2}$ bands into a lower Hubbard band (LHB) and an upper Hubbard band (UHB), opening a Mott gap as shown in Fig. 1(a). The sharp peak $\alpha$ corresponds to the optical transition from the LHB and to the UHB of the $J_{eff,1/2}$ states and the peak $\beta$ corresponds to that from the $J_{eff,3/2}$ bands to the UHB.

Notably, the two peaks show significant temperature dependence. As the temperature increases, the sharp peak $\alpha$ becomes broader and shifts to lower energy. While the heights of both peaks decrease, the SW of the peak $\alpha$ (SW$_\alpha$) increases and that of the peak $\beta$ (SW$_\beta$) decreases with increasing temperature. These spectral changes result in the blurring of the two-peak structure and a gradual decrease in the SOC-induced Mott gap. These observations are quite consistent with the resistivity behavior of Sr$_2$IrO$_4$. As temperature increases, the resistivity decreases continuously without exhibiting any discernible anomaly indicating the gradual change in the electronic structure.[16] The temperature dependence of the resistivity was found to follow the variable range hopping model[5] and the increase of the hopping rate with increasing temperature could be associated with the decrease of the optical gap.

To obtain quantitative information on the spectral changes, we fitted $\sigma(\omega)$ using the Lorentz oscillator model. As shown in Fig. 2(a), the energy of the peak $\alpha$, $\omega_\alpha$, decreases from about 0.54 eV at 10 K to 0.44 eV at 500 K. At the same time, the width of the peak $\alpha$, $\gamma_\alpha$, increases from about 0.15 eV at 10 K to 0.49 eV at 500 K. In addition to these changes, as shown in Fig. 2(b), the SW$_\alpha$ increases and the SW$_\beta$ decreases with temperature. The decrease in the SW$_\beta$ is almost the same as the increase in the SW$_\alpha$, satisfying the optical sum rule below 1.5 eV. It is also noted that $\sigma(\omega)$ shows little temperature dependence at higher energies. These experimental observations imply that the main change in electronic structure occurs in the $J_{eff,1/2}$ and $J_{eff,3/2}$ bands.

The broadening of the peaks and the SW redistribution cause the decrease in the SOC-induced Mott gap. We estimated the optical gap energy, $\Delta_{opt}$, by taking the crossing point of the line at an inflection point of the $\sigma(\omega)$ with the abscissa. As the temperature increases from 10 K to 500 K, $\Delta_{opt}$ changes from 0.41 to 0.08 eV. Note that the change in $\Delta_{opt}$, which is more than 4500 K, is much larger than that of the temperature. To get more insight, we compared our finding with the case of semiconductors, such as Ge, InAs, and Hg$_3$In$_2$Te$_6$, whose band gaps are comparable to that of Sr$_2$IrO$_4$.[17, 18] The optical band gap of the semiconductors has been known to follow the empirical Varshni relation, i.e., $E_g(T)=E_0-aT^2/(T+b)$, where $E_g$ is the energy gap, $E_0$ is the gap value at 0 K, and $a$ and $b$ are constants.[17] We fitted $\Delta_{opt}$ of Sr$_2$IrO$_4$ and found that the constant $a$, which is related to the rate of decrease in the gap with temperature, is about four to five times larger than those of the semiconductors.[17, 18] This indicates that a simple thermal effect could not be sufficient for explaining the strong temperature dependence of the Mott gap of Sr$_2$IrO$_4$.

It is worthwhile to compare the temperature-dependent evolution in $\sigma(\omega)$ of the 5$d$ Mott insulator $Sr_2IrO_4$ with those of 3$d$ Mott insulators. Previous optical studies on 3$d$ Mott insulators, such as $LaTiO_3$,[19] $YTiO_3$,[20] and $Yb_2V_2O_7$,[21] showed that their electronic structure exhibited little temperature dependence. In particular, we note that $LaTiO_3$ showed a structural transition at its magnetic transition temperature. Structural studies using x-ray and neutron diffraction techniques indicated a structural transition near the magnetic transition temperature,[22, 23] and an optical study on $LaTiO_3$ showed that the additional phonon modes appeared below the magnetic transition temperature due to the structural transition.[19] However, the electronic structure and Mott gap of $LaTiO_3$ hardly changed with temperature. The behavior of $\sigma(\omega)$ in $LaTiO_3$ is quite distinguished from that of $Sr_2IrO_4$, where $\sigma(\omega)$ of $Sr_2IrO_4$ showed strong temperature dependence without a structural transition.[24] The strong temperature dependence of $\sigma(\omega)$ without the structural transition is likely to be associated with the strong hybridization and resulting sensitivity to the electron-phonon interaction of 5$d$ TMOs due to the extended 5$d$ orbitals.

B. Theoretical calculation on electronic structure with bandwidth change

We now discuss possible origins of the significant electronic structure change in relation to the change in the bandwidth of Ir $J_{eff}$ states. The electronic bandwidth of layered perovskite is controlled by the in-plane metal-oxygen-metal bond angle, that is, the rotation of the metal-oxygen octahedron about the $c$ axis, which can induce an insulator-metal transition.[25] Generally, the active thermal fluctuation favors a high-symmetry phase of crystal structure, and hence the in-plane Ir-O-Ir bond angle can increase with temperature. Given the strong electron-phonon interaction of the 5$d$ TMOs, the change in the bond angle could lead to the large changes in the electronic structure. To simulate the electronic structure changes according to the variation in bandwidth, we preformed LDA+$U$+SOC calculation using different values of the bond angle. It was reported that the Ir-O-Ir bond angle is about 157° at 10 K.[24] We gradually increased the bond angle from 157° to 170° and obtained the corresponding density of states (DOS). Figure 3 shows the DOS of $Sr_2IrO_4$ between -1.5 and 1.0 eV where the Ir 5$d$ $t_{2g}$ orbital states are the main contributors. The DOS above the Fermi energy, $E_F$, corresponds the UHB of the $J_{eff,1/2}$ states. The DOS between -1.5 and -0.5 eV is from the $J_{eff,3/2}$ bands, and that between -0.5 and 0.0 eV (0.0 and 0.5 eV) is from the LHB (UHB) of the $J_{eff,1/2}$ states.

The calculated electronic structure changes as a function of the bond angle are qualitatively similar with the changes in the experimental $\sigma(\omega)$. As the Ir-O-Ir bond angle increases, the DOS of the $J_{eff,3/2}$ bands decreases and that of the $J_{eff,1/2}$ bands increases, indicating a DOS redistribution between the two bands. At the same time, the Mott gap between the LHB and UHB of the $J_{eff,1/2}$ states decreases. In the experimental $\sigma(\omega)$, the SW transfer from the peak $\beta$ to the peak $\alpha$ occurred and the Mott gap decreased with increasing temperature. Note that the experimental $\Delta_{opt}$ decreases by about 0.15 eV when temperature changes from 10 K to 300 K as shown in Fig. 2(c). The gap in the calculation decreases by about 0.13 eV when the bond angle changes from 157° to 170°. These results indicate that the change in bond angle might be larger than 13° to reproduce the temperature-induced change in $\Delta_{opt}$. However, a previous structural study on $Sr_2IrO_4$ showed that the bond angle increased by about 1° as temperature increased from 10 K to room temperature,[26] which is much smaller than that used in the calculation. Although the structural studies at higher temperatures are needed for direct comparison with $\sigma(\omega)$, it is hardly expected that the change in the bond angle could be large enough to induce the observed electronic structure changes. This suggests that the simple lattice distortion could not explain the large change of the bandwidth of $Sr_2IrO_4$ with temperature variation.

C. Comparison with optical spectra of $La_2CuO_4$: excitonic effect and electron-phonon coupling.

It is interesting to note that the temperature dependence of $\sigma(\omega)$ of $Sr_2IrO_4$ is quite similar with that of $La_2CuO_4$. Falck *et al*. reported the temperature-dependent optical spectra of $La_2CuO_4$.[27] A narrow charge transfer peak was observed in the optical spectra of $La_2CuO_4$. The narrowness of the charge transfer peak was attributed to the electron-hole interaction, i.e., excitonic effect, which can dramatically enhance matrix elements for interband transition in two-dimensional system.[27] As temperature increased, the charge transfer peak became broader and shifted to lower energy. At the low temperatures below 100 K, the peak energy did not show discernible change and it decreased continuously with increasing temperature above 100 K. These temperature-induced changes in $\sigma(\omega)$ were explained in terms of the coupling of charge carriers to a optical phonon mode, i.e., electron-phonon coupling. (The absence of temperature dependence below 100 K was explained in terms of the freeze-out of optical phonons.) It should be noted that the temperature dependences of the peak $\alpha$ and $\Delta_{opt}$ in $\sigma(\omega)$ of $Sr_2IrO_4$, shown in Fig. (2), are quite similar with that of the charge transfer peak of $La_2CuO_4$. The similarity of the changes in the optical spectra of $La_2CuO_4$ and $Sr_2IrO_4$ and the two-dimensional character suggest that some excitonic effect and electron-phonon coupling could play important roles for the observed changes in $\sigma(\omega)$ of $Sr_2IrO_4$.

D. Phonon dynamics

To gain some insight into the temperature-dependent lattice dynamics, we examined the phonon spectra of $Sr_2IrO_4$. Figure 4 shows the temperature-dependent phonon spectra. The phonon modes are classified into three groups: external, bending, and stretching modes.[28, 29] Figures 4(a)-4(c) show the external, bending, and stretching modes, respectively. While the external mode is related to the vibrations of the Sr ions against $IrO_6$ octahedra, the bending and stretching modes are related to the modulation of the Ir-O-Ir bond angle and the Ir-O bond length, respectively.

As shown in Figs. 4(a)-4(c), the change in the phonon modes with temperature is rather gradual and no clear split of the phonon modes is observed, indicating the absence of a structural transition.[24, 26]

To be interesting, the bending mode phonon shows the strongest temperature dependence among the observed phonon modes. Figure 4(d) shows the phonon frequency $\omega_{ph}$ of each mode normalized to $\omega_{ph}$(10 K). While the $\omega_{ph}$ of the other phonon modes changes less than 1%, that of the higher-frequency bending mode, which is related to the modulation of the in-plane bond angle,[28] changes by about 3%. (The change in the $\omega_{ph}$ is about 1 cm$^{-1}$ for the external and lower-frequency bending modes and 3 cm$^{-1}$ for the stretching mode, respectively. The change of the $\omega_{ph}$ of the higher-frequency bending mode is about 10 cm$^{-1}$.) These observations imply that the electronic structure might be coupled to the bending mode phonon.

E. Correlation between magnetic ordering and electronic structure

Finally we check the possible effects of the magnetic ordering on the electronic structure. Sr$_2$IrO$_4$ shows weak ferromagnetic ordering below $T_M$=240 K, which originates from the canted antiferromagnetic ordering of the $J_{eff}$=1/2 moments in the in-plane.[5, 16] It is noted that the SOC-induced $J_{eff,1/2}$ state is the complex state composed of the orbitals and the spins. As the electronic orbitals are locked-in by the lattice, the canting of the $J_{eff}$=1/2 moment follows the rotation of the IrO$_6$ octahedra.[11] Therefore, the magnetic property of the $J_{eff,1/2}$ state is susceptible to the local lattice change, such as the rotation of the IrO$_6$ octahedra. Conversely, a long range magnetic ordering in the $J_{eff,1/2}$ state can make the IrO$_6$ rotation stiff. In this respect, the magnetic and electronic states could be closely coupled to each other.

The temperature dependence of $\Delta_{opt}$ suggests that the electronic structure and magnetic ordering are closely coupled. As shown in Fig. 2, $\Delta_{opt}$ changes little below 200 K and its decrease become faster above 200 K which is near the magnetic transition temperature. To see the rate of the change in $\Delta_{opt}$ more clearly, we differentiated $\Delta_{opt}(T)$ with respect to temperature. As shown in the inset of Fig. 2(c), the change in $\Delta_{opt}$ is largest near the magnetic transition temperature as indicated by the gray rectangle. Optical studies under high magnetic fields could provide further information on the coupling of the electronic and magnetic structures of the $J_{eff,1/2}$ Mott state.[30]

IV. SUMMARY

We investigated the temperature-dependent optical conductivity spectra of the $J_{eff,1/2}$ Mott insulator Sr$_2$IrO$_4$. We observed significant electronic structure changes, which were associated with the decrease of the spin-orbit coupling-induced Mott gap. First-principles calculation with varied Ir-O-Ir bond angles demonstrated the change of electronic structure with the bandwidth control, which appears to be consistent with our experimental findings. We also discussed the effect of excitonic effect, electron-phonon coupling, and magnetic ordering on the electronic structure of the $J_{eff,1/2}$ state. It should be emphasized that these results exhibit characteristic features of 5*d* transition metal oxides. The extended character of 5*d* orbitals provides the strong electron-phonon coupling. Furthermore, the spin-orbit coupling presents the coupling among spin, orbital and lattice. Our study clearly demonstrates that these characteristic features make the electronic structure of the 5*d* transition metal oxide be affected by the electron-phonon interaction and by magnetic ordering.

ACKNOWLEDGEMENT

This research was supported by Basic Science Research Program through the National Reserarch Foundation of Korea funded by the Ministry of Education, Science and Technology (No. 2009-0080567), and the Korean Science and Engineering Foundation through the ARP (R17-2008-033-01000-0). The experiments at Pohang Light Source were supported in part by Ministry of Science and Technology and Pohang University of Science and Technology. YSL was supported by the Soongsil University Research Fund.

*Current address: Department of Applied Physics, Multiferroics Project, Exploratory Research for Advanced Technology, Japan Science and Technology Agency, University of Tokyo, Tokyo 113-8656, Japan

†Current address: Materials Science and Technology Division, Oak Ridge National Laboratory, Oak Ridge, TN 37831, USA.

‡ylee@ssu.ac.kr

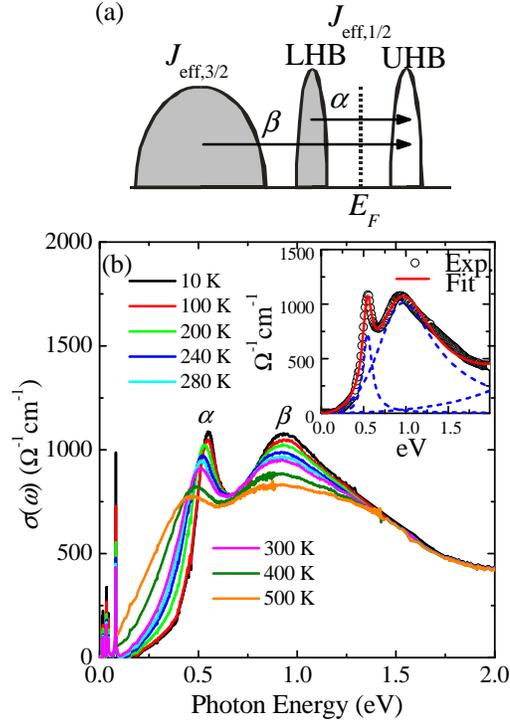

Fig. 1 (color online). (a) Schematic band diagram of the electronic structure of $Sr_2IrO_4$. $E_F$ represents the Fermi level. Peak $\alpha$ corresponds to the optical transition from the lower Hubbard band to the upper Hubbard band of the $J_{eff,1/2}$ states. Peak $\beta$ corresponds to the optical transition from the $J_{eff,3/2}$ band to the upper Hubbard band of the $J_{eff,1/2}$ states. (b) Temperature-dependent optical conductivity spectra $\sigma(\omega)$ of $Sr_2IrO_4$. As temperature increases, the peaks $\alpha$ and $\beta$ become broader and the Mott gap decreases. The sharp spikes below the optical gap energy are due to optical phonon modes. The inset shows the result of Lorentz oscillator fit for $\sigma(\omega)$ at 10 K. Open circle and red line represent experimental and fitted $\sigma(\omega)$, respectively.

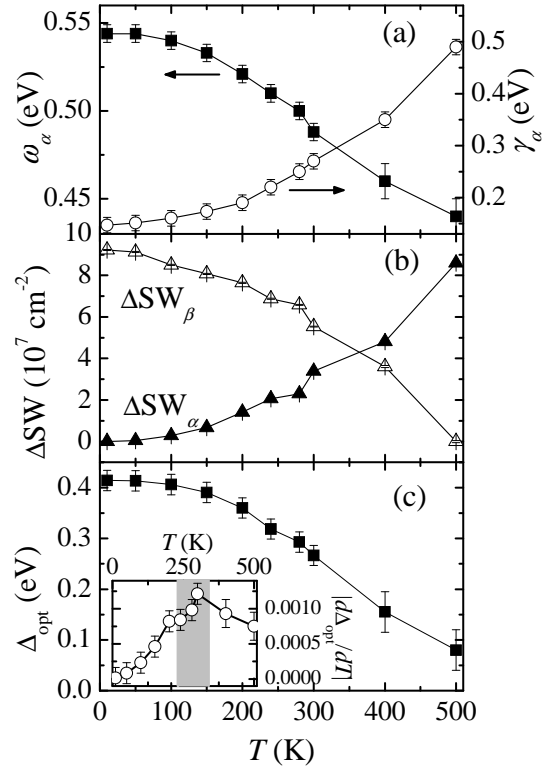

Fig. 2. (a) Temperature-dependent changes in energy (solid square) and width (open circle) of the peak $\alpha$. (b) Temperature-dependent changes in the optical spectral weight of the peaks $\alpha$ (solid triangle) and $\beta$ (open triangle). (c) Temperature-dependent change in the optical gap $\Delta_{opt}$. The inset shows the value of the gap differentiated with respect to temperature.

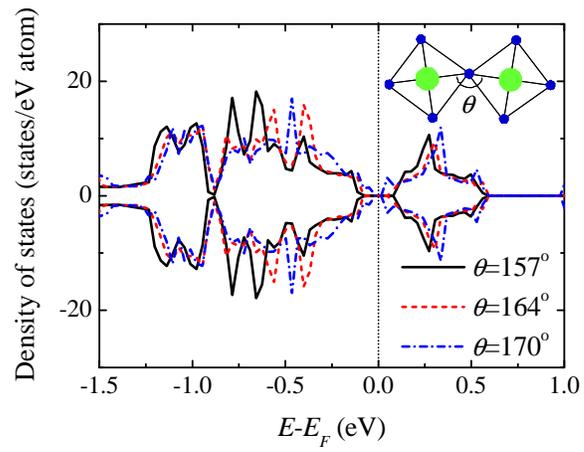

Fig. 3 (color online). Total densities of states with variation in the Ir-O-Ir bond angle. $E_F$ represents the Fermi level. The inset shows the rotation of the two neighboring octahedra. The large green and small blue circles represent Ir and O atoms, respectively. $\theta$ denotes the Ir-O-Ir bond angle.

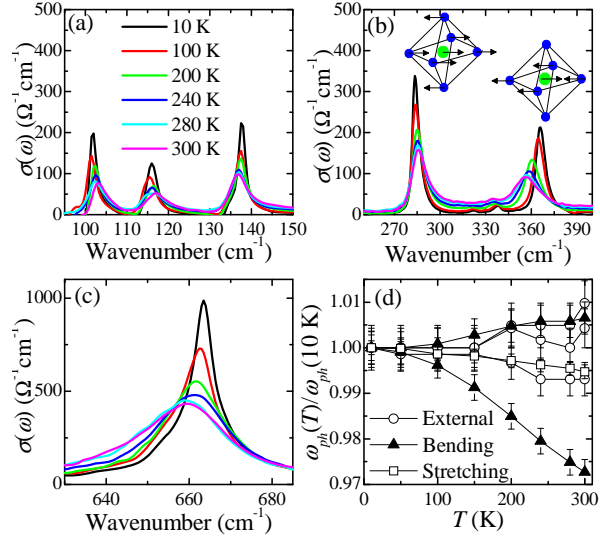

Fig. 4 (color online). Temperature-dependent phonon spectra corresponding to (a) external, (b) bending, and (c) stretching modes. The insets of (b) show the atomic displacements corresponding to the bending modes (Ref. 23). The large green and small blue circles represent Ir and O atoms, respectively. (d) Peak positions of the phonon modes $\omega_{ph}$ normalized to those at 10 K.